\documentclass[%
 reprint,
superscriptaddress,
 amsmath,amssymb,
pra,
aps,
floatfix]{revtex4-2}

\usepackage{graphicx}
\usepackage{dcolumn}
\usepackage{bm}
\usepackage{placeins}
\usepackage{xcolor}

\usepackage{array}
\usepackage{booktabs}

\usepackage{graphicx}
\usepackage{cleveref}

\newcommand{\panelref}[2]{Fig.~\ref{#1}(#2)}

\begin{document}

\preprint{APS/123-QED}

\title{Radio-frequency reflectometry in silicon carbide large-area transistors}

\author{Alexander Zotov}

\email{Email: alexander.zotov@strath.ac.uk}
\affiliation{
Department of Physics, SUPA, University of Strathclyde, Glasgow G4 0NG, United Kingdom}

\author{Conor McGeough}%
\affiliation{
Department of Physics, SUPA, University of Strathclyde, Glasgow G4 0NG, United Kingdom}
\author{Megan Powell}%

\affiliation{
Department of Physics, SUPA, University of Strathclyde, Glasgow G4 0NG, United Kingdom}

\author{Alessandro Rossi}
\email{Email: alessandro.rossi@strath.ac.uk}
\affiliation{
Department of Physics, SUPA, University of Strathclyde, Glasgow G4 0NG, United Kingdom}
\affiliation{National Physical Laboratory, Hampton Road, Teddington TW11 0LW, United Kingdom}

\date{\today}

\begin{abstract}
Radio-frequency (RF) reflectometry is widely used for high-bandwidth readout of semiconductor quantum devices at cryogenic temperatures, but its application has mainly been limited to nanoscale structures with relatively small capacitances. Here, we investigate RF readout in a different regime by applying gate-based reflectometry to a large-area silicon carbide transistor with parasitic capacitances orders of magnitude larger than those of typical quantum devices, conditions normally expected to hinder RF readout. We observe a gate-dependent RF response which degrades and eventually vanishes as temperature is lowered, although MOSFET operation in DC transport is maintained down to deep cryogenic temperatures. We attribute this behaviour to impedance changes introduced by carrier freeze-out in the transistor drift region, and propose a modified circuit configuration designed to restore sensitivity under these conditions. These results establish how parasitic pathways and device geometry can limit RF readout, providing insight into the design of scalable cryogenic-CMOS quantum systems.
\end{abstract}

\maketitle


\section{\label{sec:intro}Introduction}
Radio-frequency (RF) and microwave techniques are widely used in the field of quantum computing for the fast readout and control of quantum bits (qubits)~\cite{bardin}. In semiconductor systems, RF reflectometry enables high-bandwidth detection of small impedance changes by embedding a device within a resonant circuit and monitoring the reflected RF signal~\cite{Vigneau2022APR}. 

Recently, gate-based reflectometry has attracted significant attention as a scalable readout approach for lithographically defined quantum-dot (QD) qubit systems~\cite{colless, West2019}. In contrast to conventional architectures that rely on dedicated charge detectors, gate-based reflectometry allows the qubit gate electrode itself to act as the sensing element. This reduces device complexity and wiring overhead, which may be advantageous for scaling large arrays~\cite{Schaal2019, Ruffino2022}. Such considerations are particularly important in the context of emerging cryo-CMOS platforms, where multiplexed control and readout lines are integrated in close proximity to quantum devices~\cite{Thomas2025}. While this approach offers a pathway to large-scale integration, it also introduces additional parasitic impedances and alternative RF current pathways that can degrade readout sensitivity. Understanding how RF reflectometry behaves in the presence of large parasitic capacitances and complex circuit environments is therefore essential for scalable quantum architectures. In this context, studying large-area semiconductor devices could provide a useful testbed, as their intrinsic capacitances and multiple conductive pathways mimic the conditions expected in densely integrated cryo-CMOS systems, where signals may be redistributed through unintended circuit paths.\\\indent
Scalability is also a central motivation for developing CMOS-compatible QD platforms that can leverage mature silicon manufacturing technologies~\cite{Gonzalez-Zalba2021}. Beyond silicon, other CMOS materials are beginning to attract interest for quantum technologies~\cite{Scappucci2021, olga}. Among these, silicon carbide (SiC) has recently emerged as a promising platform. In addition to its established role in power electronics~\cite{Kimoto_rev}, SiC hosts optically addressable spin defects and has been proposed for hybrid quantum technologies~\cite{Awschalom2018,son_rev}. Extending RF readout techniques to SiC-based devices would therefore be an important step toward exploring its potential within scalable quantum architectures.\\\indent
In this work, we demonstrate gate-based RF reflectometry of a silicon carbide transistor in a regime of exceptionally large parasitic capacitances. The device studied is a vertical SiC power MOSFET with a large gate area, which would normally be expected to hinder RF readout due to parasitic capacitances between gate and the ohmic contacts providing low-impedance leakage pathways to ground~\cite{Taskinen2008RSI, Liu2021PRApp}. Nevertheless, we observe a clear gate-dependent reflectometry response at room temperature. As the temperature is reduced, this sensitivity systematically degrades despite normal MOSFET operation in DC transport. Using a lumped-element circuit model, we show that 
the loss of sensitivity at low temperature is consistent with a redistribution of RF current through parasitic circuit paths when the drift-region resistance increases due to carrier freeze-out. Finally, we propose a modified circuit configuration designed to restore sensitivity under cryogenic conditions.

\begin{figure*}[!t]
\centering
\includegraphics[width=\textwidth]{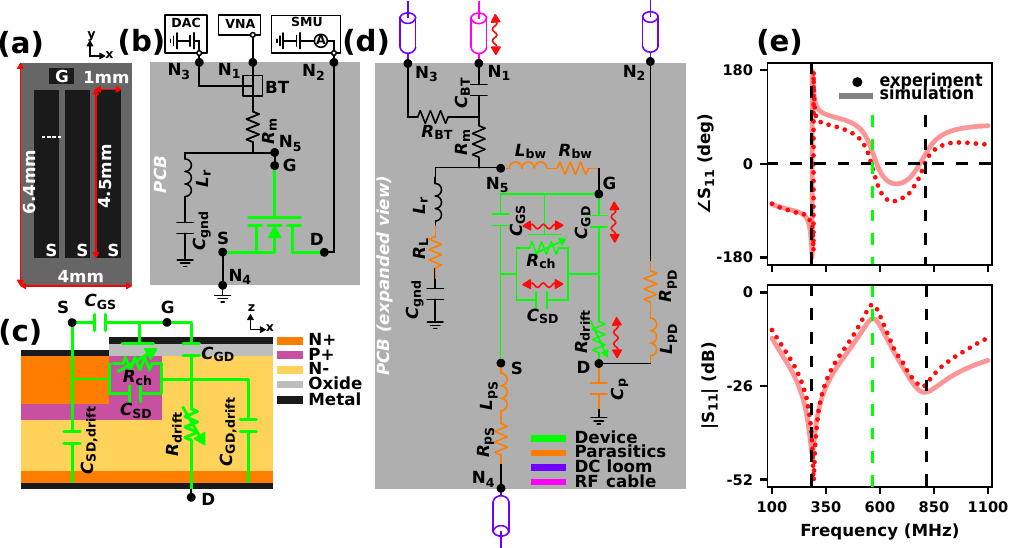}
\caption{(a) Schematic drawing (not to scale) of the bare-die commercial chip containing the DUT (Wolfspeed CPM2-1200-0025A), showing the source (S) and gate (G) metal pads on the top surface; the drain (D) contact occupies the full backside of the die and is not shown.
(b) Simplified PCB layout and measurement set-up for combined DC transport and RF reflectometry. The RF excitation and DC gate bias are combined through a bias tee (BT).
(c) Cross-sectional schematic of the vertical MOSFET along dashed white line in panel (a), with small-signal elements overlaid, indicating the physical origin of the capacitances and resistances included in the model.  $R_{\mathrm{drift}}$ is an effective resistance that includes JFET and drift resistance.
(d) Expanded lumped-element equivalent circuit model including the DUT (green), surface-mount passive components (black), and PCB/interconnect parasitics (orange). All component values are reported in Table~I of Appendix~A. The dominant RF current path through the equivalent circuit is indicated by red arrows.
(e) Room-temperature phase (top) and magnitude (bottom) of the reflection coefficient $S_{\mathrm{11}}$ measured experimentally (dots) and obtained from circuit simulations (solid lines). Dashed lines indicate the zero-crossings of the phase, with the resonant feature at $f_0=566~\mathrm{MHz}$ highlighted in green.}
\label{fig:intro}

\end{figure*}



\section{\label{sec:meth}Methods}
\subsection{Device and PCB}


The device under test (DUT) is a bare-die n-channel vertical 4H-SiC power MOSFET developed by Wolfspeed (CPM2-1200-0025A)~\cite{WolfspeedCPM2Datasheet}. The top side dimensions of the chip, PCB schematic and lumped-element model are shown in Fig.~\ref{fig:intro}. The die was glued and wire-bonded onto a custom printed circuit board (PCB) for combined DC transport and RF reflectometry. An inductor $L_{\mathrm{r}} = 8~\mathrm{nH}$ was placed in parallel with the DUT to form a resonant circuit~\cite{Vigneau2022APR}. Note that 
vertical power MOSFETs typically contain a lightly doped region between the channel and drain, which is designed to support high off-state breakdown voltages~\cite{baliga}. This is often referred to as \textit{drift region}. In our case, given that we only apply modest voltages for DC and RF characterisation, it will mainly contribute as a sizable resistive and/or capacitive pathway at cryogenic conditions. 
\subsection{Measurement set-ups}
For DC characterisation, an SMU was connected to node ${\mathrm{N_2}}$ via a constantan cryogenic loom, allowing for application of the drain-source voltage ($V_{\mathrm{DS}}$) and measurement of the drain-source current ($I_{\mathrm{DS}}$). A DAC was connected to node ${\mathrm{N_3}}$ via the same type of loom to apply the gate-source voltage ($V_{\mathrm{GS}}$).
\\\indent RF reflectometry measurements were performed in the $S_{\mathrm{11}}$ mode of a VNA connected to node ${\mathrm{N_1}}$ via RF coaxial cables, as shown in \panelref{fig:intro}{d}. 
In principle, the large gate capacitances of the vertical MOSFET, of order a few nF (see Table~I), allow RF signals applied to the gate to capacitively couple into the source and drain terminals~\cite{Taskinen2008RSI,Liu2021PRApp,Volk2019NanoLett}, so the associated parasitic signal paths must be included in the model. The circuit components and parasitic elements included in the model are listed
in Table I (Appendix~A) and depicted in Fig.~\ref{fig:intro}(d). \\\indent In general, for large-area samples, grounding strategies are particularly important. If grounding is applied at the end of the loom, RF signals can propagate along the DC wiring and introduce additional parasitic impedance contributions. By contrast, local grounding at the PCB suppresses these effects and yields a response that is well modelled by the lumped-element circuit shown in \panelref{fig:intro}{d}. A comparison of the RF response for various grounding configurations is provided in Appendix~B. To minimise additional parasitic contributions from the DC loom wiring, the drain ${\mathrm{N_2}}$ and source ${\mathrm{N_4}}$ nodes were grounded locally at the PCB DC socket, while node ${\mathrm{N_3}}$ was connected to the room-temperature breakout box through the loom (and a bias-tee) to allow application of a DC gate voltage.

To study the effects of temperature on the reflectometry readout, the PCB and DUT were thermally anchored to the mixing-chamber plate of a dilution refrigerator. As discussed in the next section, the reflectometry contrast decreases with decreasing temperature and falls below the VNA noise floor at and below $T\approx45$~K (see Appendix~C). Measurements at lower temperatures showed no meaningful gate-dependent effects. Hence, in the remainder of the manuscript, we focus on comparing and contrasting observations at room temperature and $T=28$~K, chosen as a representative cryogenic temperature of operation.

\begin{figure*}[!t]
\centering
\includegraphics[width=\textwidth]{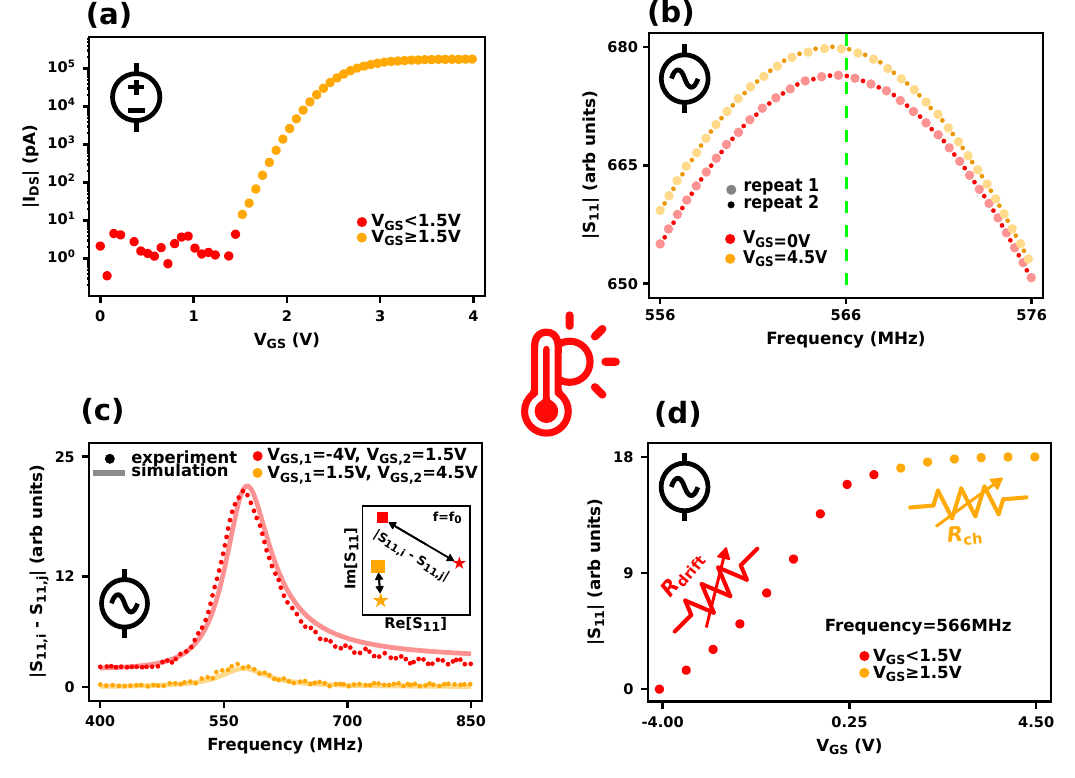}
\caption{
(a) DC transfer characteristics measured at room temperature and $V_{\mathrm{DS}} = 50\,\mu\text{V}$, showing the onset of transport through the channel at $V_{\mathrm{th}}^{\mathrm{RT}}\approx1.5~\mathrm{V}$.
(b) Magnitude of the measured reflection coefficient $|S_{11}|$ at room temperature for $V_{\mathrm{GS}}=0~\mathrm{V}$ and $V_{\mathrm{GS}}=4.5~\mathrm{V}$, including repeats.
(c) Frequency-dependent magnitude of the change in the reflection coefficient, $|\Delta S_{\mathrm{11}}|$ (with offset for clarity), probing the gate effect for voltages smaller (red) and larger (orange) than threshold voltage at room temperature. Dots show experimental data and solid lines show the corresponding circuit simulations. In the simulations, the resistive elements of the DUT varied as follows: for $V_{\mathrm{GS}}<V_{\mathrm{th}}^{\mathrm{RT}}$, $R_{\mathrm{ch}}=10~\mathrm{M}\Omega$ and $0.2~\Omega<R_{\mathrm{drift}}<0.5~\Omega$; for $V_{\mathrm{GS}}>V_{\mathrm{th}}^{\mathrm{RT}}$, $R_{\mathrm{drift}}=0.2~\Omega$ and $0.2~\Omega<R_{\mathrm{ch}}<10~\mathrm{M}\Omega$. The inset shows the complex $S_{11}$ values used to calculate $|\Delta S_{11}|$ at $f=f_0$, with the two pairs offset for clarity. 
(d) Measured magnitude of the reflection coefficient $|S_{11}|$ at $f_{0}$ as a function of gate voltage, illustrating the dominant sensitivity to changes in the effective drift-region resistance, when $V_{\mathrm{GS}}<V_{\mathrm{th}}^{\mathrm{RT}}$ (red) and the saturation of the response once the channel resistance becomes small for $V_{\mathrm{GS}}>V_{\mathrm{th}}^{\mathrm{RT}}$ (orange).}

\label{fig:hot}

\end{figure*}

\begin{figure*}[!t]
\centering
\includegraphics[width=\textwidth]{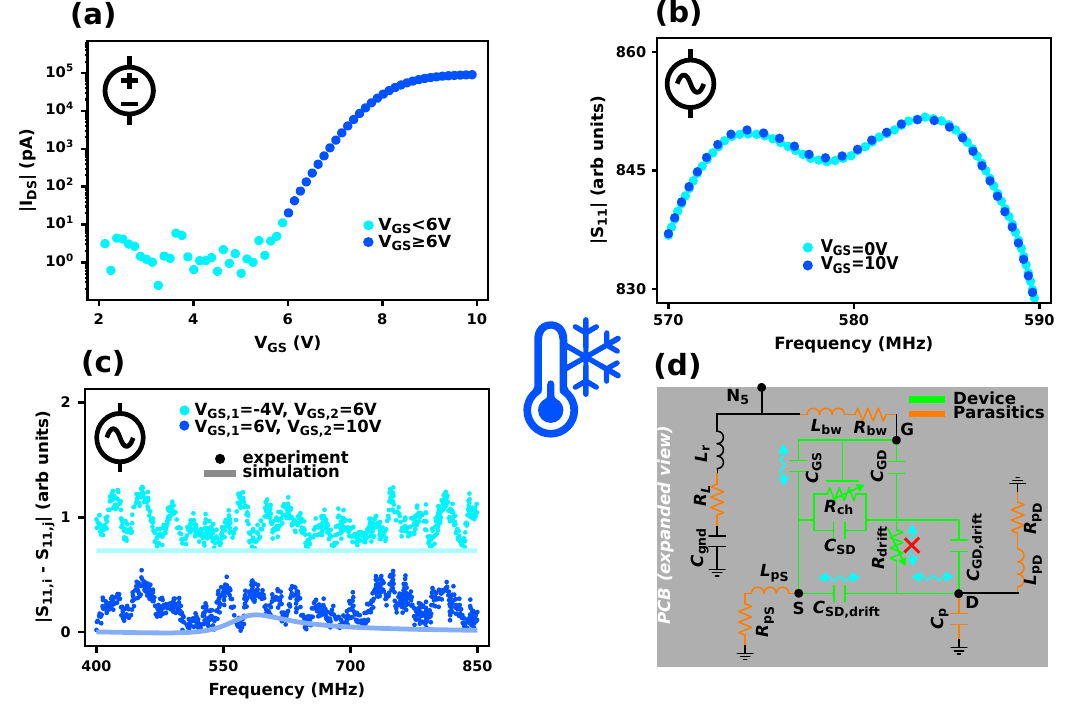}
\caption{(a) DC transfer characteristic measured at $T=28~\mathrm{K}$ and $V_{\mathrm{DS}}=10$~mV, showing gate-controlled turn-on and $V_{\mathrm{th}}^{\mathrm{cryo}}\approx6$~V.
(b) Magnitude of the reflection coefficient for $V_{\mathrm{GS}}=0~\mathrm{V}$ and $V_{\mathrm{GS}}=10~\mathrm{V}$,  $T=28~\mathrm{K}$.
(c) Frequency dependent magnitude of the change in the reflection coefficient, $|\Delta S_{11}|$ (with offset for clarity), probing the gate effect for voltages smaller (light blue) and larger (dark blue) than threshold voltage at $T=28~\mathrm{K}$. Dots show experimental data and solid lines show the corresponding circuit simulations. In the simulations, the resistive elements of the DUT varied as follows: for $V_{\mathrm{GS}}<V_{\mathrm{th}}^{\mathrm{cryo}}$, $R_{\mathrm{ch}}=10~\mathrm{M}\Omega$ and $3.4~\mathrm{k}\Omega<R_{\mathrm{drift}}<8.6~\mathrm{k}\Omega$; for $V_{\mathrm{GS}}>V_{\mathrm{th}}^{\mathrm{cryo}}$, $R_{\mathrm{drift}}=3.4~\mathrm{k}\Omega$ and $0.2~\Omega<R_{\mathrm{ch}}<10~\mathrm{M}\Omega$. Calculation for $R_{\mathrm{drift}}=8.6~\mathrm{k}\Omega$ is shown in Appendix D, while the lower value $R_{\mathrm{drift}}=3.4~\mathrm{k}\Omega$ is obtained by applying the same relative gate-induced reduction used for the room-temperature drift resistance.
(d) Expanded circuit model illustrating the dominant RF current paths through the device via $C_{\mathrm{SD,drift}}$ and $C_{\mathrm{GS}}$ at low temperature.}
\label{fig:cold}

\end{figure*}



\section{\label{sec:res}Results}
\subsection{Room temperature}
In \panelref{fig:hot}{a}, we show the room-temperature DC transfer characteristic of the device, measured with a constant drain-source bias. We define the threshold voltage, $V_{\mathrm{th}}$, as the gate voltage at which $|I_{\mathrm{DS}}|>5\sigma_{I_{\mathrm{DS}}}$. Here $\sigma_{I_{\mathrm{DS}}}$ is the standard deviation of the drain-source current measured with the DUT in the non-conducting state. According to this definition, the device exhibits $V_{\mathrm{th}}^{\mathrm{RT}} \approx 1.5~\mathrm{V}$ at room temperature $(5\sigma_{I_{\mathrm{DS}}}^{\mathrm{RT}}=12~\mathrm{pA})$. The measured on-resistance, $R^\textup{RT}_{\mathrm{DS,on}}$, is consistent with the manufacturer's datasheet at sub-$\Omega$ levels~\cite{WolfspeedCPM2Datasheet}.\\\indent
We now turn to discuss the reflectometry readout. In \panelref{fig:intro}{e} we show the measured $S_{\mathrm{11}}$ response which exhibits two resonant features associated with the DUT at approximately 566~MHz and 790~MHz, which appear only when the device is connected. An additional resonant feature at approximately 290~MHz arises from the bias-tee capacitor and resonator inductor $L_{\mathrm{r}}$ and does not involve the DUT (see Appendix~B). The higher-frequency resonances arise from the interaction of $L_{\mathrm{r}}$ with the effective impedance of the gate-source and gate-drain paths on the PCB, including parasitic elements ($L_{\mathrm{pS}}$, $R_{\mathrm{pS}}$, $L_{\mathrm{pD}}$, and $R_{\mathrm{pD}}$). RF reflectometry measurements were performed in the vicinity of the resonant feature at approximately 566~MHz. This frequency range was chosen as it exhibits the strongest sensitivity to gate-induced changes in the device impedance. In the remainder of this work we denote the frequency associated with this feature by $f_{\mathrm{0}}$.\\\indent 
In \panelref{fig:hot}{b} we show the measured magnitude of the reflection coefficient $|S_{11}|$ in the vicinity of $f_{0}$ for two values of \(V_{\mathrm{GS}}\), chosen such that one is smaller and one is larger than \(V_{\mathrm{th}}^{\mathrm{RT}}\) in order to probe channel effects. Repeated measurements at each gate voltage suggest that a reproducible change in the magnitude of $|S_{11}|$ is observed between the two bias points, demonstrating that the reflectometry response is sensitive to gate-induced changes in the effective RF impedance of the device. The repeatability of the measurements indicates that the observed signal shift is not attributable to device instability.

To determine the frequency at which the gate-induced shift is maximised, we evaluate the magnitude of the change in the complex reflection coefficient, $|\Delta S_{\mathrm{11}}|=|S_{\mathrm{11,i}}-S_{\mathrm{11,j}}|$, between pairs of gate voltages in selected intervals. The first interval is for the transistor in the OFF state and corresponds to \(-4~\mathrm{V} < V_{\mathrm{GS}} < V_{\mathrm{th}}^{\mathrm{RT}}\), while the second corresponds to \(V_{\mathrm{th}}^{\mathrm{RT}} < V_{\mathrm{GS}} < 4.5~\mathrm{V}\), i.e., transistor in the ON state. The resulting frequency dependence of $|\Delta S_{\mathrm{11}}|$ is shown in \panelref{fig:hot}{c}, together with the corresponding simulated response obtained from our lumped-element circuit model. For both gate-voltage intervals, the maximum change in $|\Delta S_{\mathrm{11}}|$ occurs near $f_{\mathrm{0}}$, in agreement with the model prediction. 

Notably, the magnitude of $|\Delta S_{11}|$ is substantially larger for gate voltages $V_{\mathrm{GS}} < V_{\mathrm{th}}^{\mathrm{RT}}$ than for $V_{\mathrm{GS}} > V_{\mathrm{th}}^{\mathrm{RT}}$. This indicates that the dominant contribution to the reflectometry response arises below threshold. In other words, a measurable RF response is present even in the absence of a fully formed conducting transistor channel. This behaviour is reproduced in the circuit model by a gate-dependent resistance in the current path to the drain electrode. We argue that this is consistent with a gate-dependent change in the resistance of the transistor drift region ($R_{\mathrm{drift}}$). The drift region of a vertical power MOSFET is indeed expected to present a bias-dependent depletion width ~\cite{baliga}. Additionally, we note that the observed RF response cannot be attributed to gate-induced capacitance changes alone: $C_{\mathrm{GD}}$ and $C_{\mathrm{GS}}$ are much larger than the channel capacitance and appear in parallel, thereby dominating the RF pathway. This makes a channel-capacitance mechanism insufficient to account for the measured contrast. 



To further evidence the gate-voltage dependence of the reflectometry response, \panelref{fig:hot}{d} shows the measured magnitude of $|S_{\mathrm{11}}|$ at $f_{\mathrm{0}}$ as a function of $V_{\mathrm{GS}}$ over the full bias range used in \panelref{fig:hot}{c}. A pronounced gating effect in $|S_{\mathrm{11}}|$ is observed at $V_{\mathrm{GS}}<V_{\mathrm{th}}^{\mathrm{RT}}$, followed by a gradual saturation at higher $V_{\mathrm{GS}}$. This is consistent with the RF response probing changes in the resistance of the drift region due to gate-induced depletion effects in the transistor sub-threshold regime~\cite{baliga}.


\subsection{Cryogenic temperature}

In \panelref{fig:cold}{a} we show the DC transfer characteristic of the device measured at a temperature of $T=28~\mathrm{K}$. 
The device exhibits a clear gate-controlled turn-on and saturation behaviour, confirming MOSFET operation under cryogenic conditions. We extract a threshold voltage at this temperature of $V_{\mathrm{th}}^{\mathrm{cryo}} \approx 6~\mathrm{V}$. This is a significant increase with respect to room temperature, and it is consistent with reduced intrinsic carrier density and charge trapping effects at low temperature ~\cite{Tian2019TED,Powell_2026}. From the saturated drain current of approximately $90~\mathrm{nA}$ at $V_{\mathrm{DS}} = 10~\mathrm{mV}$, we calculate $R_{\mathrm{DS,on}}^{\mathrm{cryo}} \approx 110~\mathrm{k}\Omega$, which is also a significant increase with respect to the room temperature value. Carrier freeze-out in the lightly doped drift region as well as Schottky effect in the contact regions are expected to be major contributors to this resistance~\cite{Liu2020LowRon1200V} (see Appendix~D). 
\\\indent In \panelref{fig:cold}{b} we show the measured magnitude of the reflection coefficient $|S_{11}|$ near the same resonant feature used in the room-temperature measurements for two gate voltages, one smaller and one larger than the cryogenic threshold. Note that the resonant feature which was located at $f_0\approx 566~$MHz at room temperature, shifts to higher frequency at low temperature (by roughly 14~MHz). This is well captured by lumped-element simulations when $R_{\mathrm{drift}}$ is increased to the $\mathrm{k}\Omega$ range, consistent with the drift-region freeze-out estimate presented in Appendix~D. The shift is attributed to a change in the effective RF pathway, which modifies the reactance and therefore the resonant frequency.  In contrast to the room-temperature behaviour, no appreciable change in $|S_{\mathrm{11}}|$ is observed between the two selected voltages. Note that the same vertical scale is used in both \panelref{fig:hot}{b} and \panelref{fig:cold}{b}. The different response indicates that, at this lower temperature, the reflectometry signal is insensitive to changes due to gate voltage over the range probed.

To quantify this behaviour, \panelref{fig:cold}{c} shows the magnitude of the change in the complex reflection coefficient, $|\Delta S_{\mathrm{11}}|$, evaluated between two gate-voltage intervals probing channel inversion. At $T=28~\mathrm{K}$, $|\Delta S_{\mathrm{11}}|$ remains below the experimental noise floor throughout the measured frequency range for both gate-voltage intervals. This absence of contrast is reproduced by the circuit model
~of \panelref{fig:cold}{d} where we show the corresponding dominant RF current paths at low temperature. At low temperature, the central effect is the increase of $R_{\mathrm{drift}}$ into the $\mathrm{k}\Omega$ range, as estimated in Appendix~D. This suppresses RF current through the gate-dependent resistive path of the DUT and redirects the dominant RF oscillation through lower-impedance parasitic pathways in the circuit. Hence, sensitivity to gate-to-channel capacitance is also not present, as the RF current flows predominantly through the lower-impedance capacitive pathway formed by $C_{\mathrm{GS}}$ and $C_{\mathrm{SD,drift}}$, which do not depend strongly on $V_{\mathrm{GS}}$. Note that drift-region capacitances, omitted from \panelref{fig:intro}{d} in the room temperature equivalent circuit, become comparatively more important in the low-temperature circuit due to the larger $R_{\mathrm{drift}}$ associated with carrier freeze-out.



\section{\label{sec:dis}Discussion}

\begin{figure}[!t]
\centering
\includegraphics[width=\columnwidth]{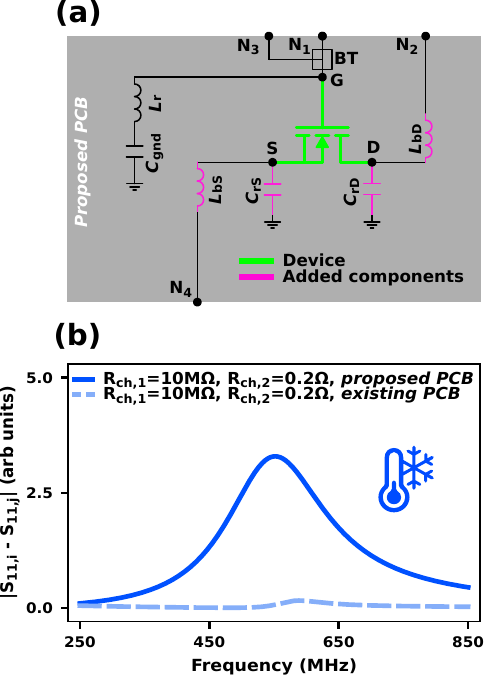}
\caption{ 
(a) Schematic of the modified PCB configuration, in which additional components (pink) are introduced to redistribute the RF current through gate-dependent DUT impedances. At each of the source and drain nodes, a capacitor provides a resonant RF path with $L_r$, while an inductor acts as an RF choke. 
(b) Simulated magnitude of the change in the reflection coefficient, $|S_{\mathrm{11,i}}-S_{\mathrm{11,j}}|$, for the original PCB configuration (dashed line) and the proposed configuration (solid line), evaluated for a change in channel resistance $0.2~\Omega < R_{\mathrm{ch}} < 10~\mathrm{M}\Omega$. Other parameters used in the simulations are $L_\mathrm{r} = 4~\mathrm{nH}$, $C_{\mathrm{rS}}=C_{\mathrm{rD}}=11~\mathrm{pF}$, $L_{\mathrm{bS}}=L_{\mathrm{bD}}=500~\mathrm{nH}$.}

\label{fig:solution}

\end{figure}

\subsection{Implications of device architecture on RF readout}
The defining features of the device studied here are its large gate area and its vertical architecture. The former gives rise to substantial parasitic capacitances and consequently low-impedance RF pathways between the gate and ohmic regions. The vertical structure incorporates a thick, lightly doped drift region that in power applications is used to sustain high electric fields and enable high-voltage operation~\cite{baliga}. In our case, this region introduces a substantial, temperature-dependent resistance and associated capacitive coupling, which strongly influence reflectometry sensitivity.
Unlike gate-based reflectometry of nanoscale devices, where sensitivity is typically dominated by changes in capacitance~\cite{Taskinen2008RSI,Liu2021PRApp}, the large-gate geometry of the vertical MOSFET studied here shifts the dominant sensitivity toward changes in effective resistance. In this device, the capacitance to the drift region is substantially larger than the capacitance between the gate and channel and appears in parallel with it, thereby diverting the RF current toward the drift region. As a result, variations in the channel capacitance produce a negligible change in the total RF impedance seen by the resonator.


At room and moderately low temperatures, our measurements indicate sensitivity to resistive changes in the drift region due to gating and/or temperature. This is notable as, to the best of our knowledge, vertical MOSFETs have not previously been studied using RF reflectometry. Although this sensitivity is inferred at the circuit level rather than extracted from a microscopic transport model, the agreement between experiment and simulation suggests that RF reflectometry can probe modifications of $R_{\mathrm{drift}}$ in vertical devices. Establishing the microscopic origin of this resistance modulation is beyond the scope of the present work.
\subsection{Response to resistive changes}
To understand how the reflectometry response varies with frequency and under cryogenic operation, it is useful to consider the sensitivity of $S_{\mathrm{11}}$ to variations in a resistive element $R_x$ in the circuit~\cite{senseRF}. For a small change $\Delta R_x$, over which the current in the circuit remains approximately constant,

\begin{equation}
\left|\Delta S_{11}\right|
\approx
\left|
\frac{2 Z_0}{\left(Z_{\mathrm{N}_1}+Z_0\right)^2}
\right|
\left|
\frac{I_x}{I_{\mathrm{N}_{\mathrm{1}}}}
\right|^2
\left|\Delta R_x\right|,
\end{equation}

where $Z_{\mathrm{{\mathrm{N_1}}}}$ is the input impedance seen from node $\mathrm{N}_{\mathrm{1}}$, $Z_{\mathrm{0}}$ is the line characteristic impedance, $I_{\mathrm{{\mathrm{N_1}}}}$ is the net RF current at the input node $\mathrm{N}_{\mathrm{1}}$, and $I_x$ is the RF current flowing through $R_x$. In the present circuit, $R_x$ corresponds to either $R_{\mathrm{ch}}$ or $R_{\mathrm{drift}}$. The sensitivity is therefore determined both by the prefactor $\left|2Z_0/\left(Z_{\mathrm{N}_{\mathrm{1}}}+Z_0\right)^2\right|$, which is largest for small $Z_{\mathrm{{\mathrm{N_1}}}}$, and by the current-participation factor $\left|I_x/I_{\mathrm{N}_{\mathrm{1}}}\right|^2$, which is largest when the current through the relevant device resistance is maximised relative to the current at the input node. This framework explains the enhanced reflectometry response near $f_0$ observed in \panelref{fig:hot}{c} relative to the other resonant features. Our circuit model reveals that the input impedance near $f_0$ is $Z_\mathrm{N_1}\approx120~\Omega$. Although this yields a smaller prefactor than at the other resonant features, where $Z_\mathrm{N_1}\approx50~\Omega$, the RF current through $R_{\mathrm{drift}}$, and to a lesser extent through $R_{\mathrm{ch}}$ (when channel is near threshold), is much larger relative to $I_\mathrm{N_1}$. This larger current-participation factor produces the largest overall response.\\\indent
At cryogenic temperature, the input impedance near the relevant resonant feature remains approximately unchanged, so the prefactor and the input current $I_{\mathrm{N}_{\mathrm{1}}}$ do not account for the loss of response. Instead, the increase in the effective drift-region resistance reduces the RF current flowing through it, thereby suppressing the current-participation factor $\left|I_{\mathrm{drift}}/I_{\mathrm{N}_{\mathrm{1}}}\right|^2$. The current through $R_{\mathrm{ch}}$ is also reduced, so variations in either $R_{\mathrm{drift}}$ or $R_{\mathrm{ch}}$ produce only a small change in $S_{11}$. Under these conditions, the dominant RF current is redistributed through the lower-impedance, gate-independent parasitic capacitances, as shown in \panelref{fig:cold}{d}. This behaviour highlights that reflectometry sensitivity depends on the relative impedances of all available RF current paths in the device-circuit system.
\subsection{Proposed RF circuit}
To restore sensitivity under cryogenic conditions, we propose a modified circuit designed to satisfy two requirements. First, the dominant resonant current should include the gate-dependent channel path, so that changes in $R_{\mathrm{ch}}$ remain detectable when $R_{\mathrm{drift}}$ is large. Second, RF leakage into the external DC wiring should be suppressed, while preserving DC access to the source and drain terminals. The proposed circuit is shown in \panelref{fig:solution}{a}. A capacitor is added at the source terminal to place the gate-source-channel path within the resonant circuit formed with the resonator inductance. This increases the current participation of the channel and therefore enhances sensitivity to variations in $R_{\mathrm{ch}}$ in the cryogenic regime. A second capacitor is added at the drain terminal to provide a controlled RF path through the drift region, preserving sensitivity to variations in $R_{\mathrm{drift}}$ at room temperature.

Additional inductors are introduced at the source and drain terminals to act as RF chokes. Although the RF excitation is applied to the gate, the large gate-source and gate-drain capacitances couple RF signals strongly into the source and drain nodes. The chokes therefore present a high impedance to RF signal, suppressing leakage into the external DC wiring and confining the oscillatory current to the intended pathways. We choose to apply the RF excitation to the gate rather than to an ohmic contact, as commonly used in charge-detector reflectometry~\cite{Schoelkopf1998,Reilly2007}, because for this large-area device the gate input provides a larger impedance differential and, therefore, a more favourable option for reflectometry readout, as discussed in Appendix~E.

In \panelref{fig:solution}{b}, we compare the simulated reflectometry response of the original circuit with that of the proposed configuration for the same change in $R_{\mathrm{ch}}$. In the original circuit, the response is dominated by RF current paths through gate-independent parasitic capacitances, resulting in negligible sensitivity to $R_{\mathrm{ch}}$. In contrast, the proposed circuit exhibits a pronounced response to the same change in $R_{\mathrm{ch}}$, indicating that the added elements redistribute the RF current such that gate-dependent device impedances contribute directly to the reflectometry signal.

\section{\label{sec:conc}Conclusion and Outlook}

In this work we have demonstrated gate-based RF reflectometry of a large-area SiC vertical MOSFET. Despite the large parasitics associated with the device geometry, a gate-dependent RF response is reported. Circuit modelling shows that the reflectometry signal is primarily sensitive to variations in the effective channel and drift-region resistances rather than to capacitance changes. At cryogenic temperature, however, the reflectometry signal degrades although the MOSFET remains operational in DC transport. Our analysis indicates that this behaviour arises from architectural features specific to vertical power MOSFETs, including the lightly doped drift region, whose resistance increases due to carrier freeze-out, and the large device capacitances that divert RF current through parasitic pathways. 

More generally, these results highlight that the sensitivity of RF reflectometry is governed by the distribution of RF current within the combined device–circuit system. In this context, the large-area device studied here provides a useful proxy for the complex impedance environments expected in cryogenic-CMOS architectures, where multiplexed interconnects and control circuitry introduce additional parasitic pathways. The observed loss of sensitivity therefore illustrates a potential limitation for scalable RF readout in such systems.


\begin{acknowledgments}
We wish to acknowledge useful discussions with E. Parry and J.D. Fletcher, as well as technical support from A. Robbins. AR acknowledges support from the UKRI Future Leaders Fellowship Scheme (Grant agreement: UKRI1071).
\end{acknowledgments}

\appendix

\section{Estimation of PCB and Interconnect Parasitic Elements}
\label{app:parasitics}

We present calculations used to estimate both the parasitic circuit elements and the device parameters included in the lumped-element circuit model shown in \panelref{fig:intro}{d}. These estimates are based on device and PCB geometry, material properties, and manufacturer's datasheets.

\subsection{Bond-wire inductance}
\label{app:bondwire_L}

The self-inductance of a straight bond wire is approximated by ~\cite{Terman1943}:
\begin{equation}
L_{\mathrm{bw}} \approx \frac{\mu_0 l}{2\pi}
\left[
\ln\!\left(\frac{2l}{r}\right) - 1
\right],
\end{equation}
where $l$ is the wire length, $r$ is the wire radius and $\mu_0$ is the permeability of free space. Our aluminium bond wires have a length of
$l \approx 5~\mathrm{mm}$ and diameter $50~\mu\mathrm{m}$ ($r = 25~\mu\mathrm{m}$), hence $L_{\mathrm{bw}} \approx 5~\mathrm{nH}$.

\begin{figure}[!b]
  \centering
  \includegraphics[width=\columnwidth]{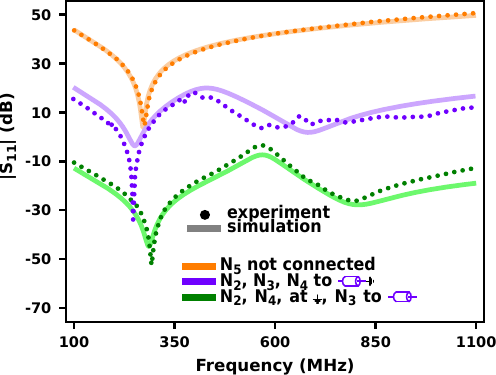}
  \caption{Measured magnitude of the reflection coefficient $|S_{11}|$ at room temperature (dots) for different grounding configurations tested (color coded), and associated circuit simulations (solid lines). Traces are shifted vertically for clarity.}
  \label{fig:s11GND}
\end{figure}

\subsection{Gate resistance and bond-wire resistance}
\label{app:bondwire_R}

The DC resistance of a single aluminum bond wire is estimated from
\begin{equation}
R_{\mathrm{bw,dc}} = \frac{\rho l}{\pi r^2},
\end{equation}
where the resistivity of aluminum is $\rho = 2.7 \times 10^{-8}~\Omega\mathrm{m}$ at room temperature. Substituting
$l = 5~\mathrm{mm}$ and $r = 25~\mu\mathrm{m}$ gives~$R_{\mathrm{bw,dc}} \approx 0.07~\Omega$.

At RF frequencies, the resistance is increased by skin effect. The skin depth is
\begin{equation}
\delta = \sqrt{\frac{2\rho}{\omega \mu_0}},
\end{equation}
with $\omega = 2\pi f$. At $f = 566~\mathrm{MHz}$, this yields $\delta \approx 3.5~\mu\mathrm{m}$.
For $r \gg \delta$, the effective cross-sectional area is reduced, and the RF resistance scales approximately as
\begin{equation}
\frac{R_{\mathrm{bw,rf}}}{R_{\mathrm{bw,dc}}} \approx \frac{r}{2\delta},
\end{equation}
giving~$R_{\mathrm{bw,rf}} \approx 0.25~\Omega$.

In addition to the bond-wire resistance, the gate path includes the intrinsic gate resistance of the MOSFET, specified by the manufacturer's datasheet as $R_{\mathrm{G,int}} = 1.1~\Omega$. In the circuit model, these contributions are combined into a single effective gate resistance $R_{\mathrm{bw}}$, which is treated as an effective series resistance. The remaining excess resistance is attributed to bond-foot contact resistance. The same resistance value is used at room and cryogenic temperature, since the reduction expected from the lower resistivity at cryogenic temperature does not materially affect the results.

\subsection{Source and drain parasitic impedances to ground}
\label{app:source_drain}

The parasitic inductances and resistances associated with the source and drain connections to ground
($L_{\mathrm{pS}}$, $R_{\mathrm{pS}}$, $L_{\mathrm{pD}}$, and $R_{\mathrm{pD}}$) include contributions from the corresponding bond wires and the PCB DC traces between the device pads and the point at which a low-impedance RF ground is established. Using a typical PCB trace inductance of order $1~\mathrm{nH/mm}$ and measured trace lengths of approximately $10$--$15~\mathrm{mm}$ yields inductance values reported in Table I.

\subsection{Parasitic capacitance between drain island and ground plane}
\label{app:cp}

The parasitic capacitance $C_\mathrm{p}$ between the drain-side metal island and the PCB ground plane is estimated using a parallel-plate capacitor model,
\begin{equation}
C_\mathrm{p} = \varepsilon_0 \varepsilon_r \frac{A}{d},
\end{equation}
where $A = 6~\mathrm{mm} \times 12~\mathrm{mm}$ is the area of the metal island, $d = 0.5~\mathrm{mm}$ is the PCB thickness, and $\varepsilon_r = 3.5$ is the relative permittivity of the Rogers~4350B substrate. Substitution gives
$C_\mathrm{p} \approx 4~\mathrm{pF}$.

In the grounded configuration used for the reflectometry measurements, this capacitance appears in parallel with additional capacitance associated with multiple DC traces connected to the same node. Although these traces are nominally grounded via a shorting plug, the final connection to the PCB ground plane has a finite impedance, such that the traces are not held at an ideal RF ground along their entire length. Each DC trace therefore behaves as an open-ended stub that contributes a capacitive admittance at the node. The combined effect of these parallel stubs increases the effective shunt capacitance beyond the simple parallel-plate estimate, and this total contribution is treated as an effective capacitance in the circuit model.

\subsection{Estimation of device capacitances and resistances from datasheet}

The intrinsic device capacitances $C_{\mathrm{GS}}$, $C_{\mathrm{GD}}$, and $C_{\mathrm{SD}}$ are estimated from the manufacturer's datasheet from the capacitance-voltage characteristics measured at $V_{\mathrm{GS}} = 0$. The values are taken at $V_{\mathrm{DS}} = 0$.

The capacitances associated with the drift region, $C_{\mathrm{GD,drift}}$ and $C_{\mathrm{SD,drift}}$, are estimated from the same dataset in the high $V_{\mathrm{DS}}$ regime where the capacitance saturates. In this regime, the depletion region extends across the full drift layer, and the measured capacitance reflects the residual capacitive coupling through the fully depleted region.

This high-$V_{\mathrm{DS}}$ condition provides an estimate of the effective capacitance when the drift region is non-conductive, which is comparable to the situation at cryogenic temperature where carrier freeze-out suppresses conduction in the drift region.

\begin{table*}[t]
\centering
\caption{Components of the lumped-element circuit model and their respective values used in the simulations.}
\label{tab:circuit_parameters}
\small
\setlength{\tabcolsep}{3pt}

\renewcommand{\arraystretch}{2}

\begin{tabular*}{\textwidth}{@{\extracolsep{\fill}}llll}
\toprule
\multicolumn{1}{l}{Component} &
\multicolumn{1}{c}{Description} &
\multicolumn{1}{l}{Value} &
\multicolumn{1}{c}{Justification} \\
\midrule

$L_{\mathrm{r}}$ & \parbox[t]{5.6cm}{Resonator inductor forming the LC circuit} & 8 nH &
\parbox[t]{3.6cm}{Datasheet nominal value} \\

$R_{\mathrm{L}}$ & \parbox[t]{5.6cm}{Effective resistance in series with $L_{\mathrm{r}}$ }
& 2.5 $\Omega$ &
\parbox[t]{3.6cm}{Datasheet nominal value} \\

$R_{\mathrm{m}}$ & \parbox[t]{5.6cm}{Series matching resistor used to enhance visibility of the resonance} & 47 $\Omega$ &
\parbox[t]{3.6cm}{Datasheet nominal value} \\

$C_{\mathrm{gnd}}$ & \parbox[t]{5.6cm}{Capacitor used to short $L_{\mathrm{r}}$ to ground for RF while blocking DC} & 10 nF &
\parbox[t]{3.6cm}{Datasheet nominal value} \\

$C_{\mathrm{BT}}$ & \parbox[t]{5.6cm}{Bias-tee capacitor coupling RF signal to the gate while blocking DC} & 47 pF &
\parbox[t]{3.6cm}{Datasheet nominal value} \\

$R_{\mathrm{BT}}$ & \parbox[t]{5.6cm}{Bias-tee feedthrough resistor providing DC gate bias} & 100 k$\Omega$ &
\parbox[t]{3.6cm}{Datasheet nominal value} \\

$L_{\mathrm{bw}}$ & \parbox[t]{5.6cm}{Bond-wire inductance between device and PCB} & 5 nH &
\parbox[t]{3.6cm}{Geometry/material estimate} \\

$R_{\mathrm{bw}}$ & \parbox[t]{5.6cm}{Effective series resistance of bond-wire and contacts} & 8 $\Omega$ &
\parbox[t]{3.6cm}{Geometry/material estimate} \\

$L_{\mathrm{pS}}$ & \parbox[t]{5.6cm}{Source bond-wire and PCB trace inductance to ground} & 14 nH &
\parbox[t]{3.6cm}{Geometry/material estimate} \\

$R_{\mathrm{pS}}$ & \parbox[t]{5.6cm}{Source parasitic resistance to ground} & 0.5 $\Omega$ &
\parbox[t]{3.6cm}{Geometry/material estimate} \\

$L_{\mathrm{pD}}$ & \parbox[t]{5.6cm}{Drain bond-wire and PCB trace inductance to ground} & 12 nH &
\parbox[t]{3.6cm}{Geometry/material estimate} \\

$R_{\mathrm{pD}}$ & \parbox[t]{5.6cm}{Drain parasitic resistance to ground} & 0.5 $\Omega$ &
\parbox[t]{3.6cm}{Geometry/material estimate} \\

$C_{\mathrm{p}}$ & \parbox[t]{5.6cm}{Parasitic capacitance between drain metal island and PCB ground plane} & 18 pF &
\parbox[t]{3.6cm}{Geometry/material estimate} \\

$C_{\mathrm{GS}}$ & \parbox[t]{5.6cm}{Gate-source capacitance of the MOSFET} & 2 nF &
\parbox[t]{3.6cm}{Device datasheet estimate} \\

$C_{\mathrm{GD}}$ & \parbox[t]{5.6cm}{Gate-drain capacitance} & 2 nF &
\parbox[t]{3.6cm}{Device datasheet estimate} \\

$C_{\mathrm{SD}}$ & \parbox[t]{5.6cm}{Source-drain capacitance} & 2 nF &
\parbox[t]{3.6cm}{Device datasheet estimate} \\

$C_{\mathrm{GD,drift}}$ & \parbox[t]{5.6cm}{Gate-drain capacitance through drift region} & 15 pF &
\parbox[t]{3.6cm}{Device datasheet estimate} \\

$C_{\mathrm{SD,drift}}$ & \parbox[t]{5.6cm}{Source-drain capacitance through drift region} & 200 pF &
\parbox[t]{3.6cm}{Device datasheet estimate} \\

$R_{\mathrm{ch}}$ & \parbox[t]{5.6cm}{Channel resistance (gate dependent)} & $10~\mathrm{M}\Omega$ to $0.2~\Omega$ &
\parbox[t]{3.6cm}{Device datasheet estimate} \\

$R_{\mathrm{drift}}$ at room temperature & \parbox[t]{5.6cm}{Effective drift-region resistance at room temperature (gate dependent)} & $0.5~\Omega$ to $0.2~\Omega$ &
\parbox[t]{3.6cm}{Device datasheet estimate} \\

$R_{\mathrm{drift}}$ at 28 K & \parbox[t]{5.6cm}{Effective drift-region resistance at cryogenic temperature (gate dependent)} & $8.6~\mathrm{k}\Omega$ to $3.4~\mathrm{k}\Omega$ &
\parbox[t]{3.6cm}{Estimated/Calculated, see Appendix D} \\

\bottomrule
\end{tabular*}
\end{table*}


\section{Effect of grounding configurations on reflection coefficient}
\label{app:reflection}

In Fig.~\ref{fig:s11GND}, we show the measured room-temperature reflection coefficient $|S_{\mathrm{11}}|$ as a function of frequency for different circuit configurations, together with simulated responses obtained from the corresponding lumped-element circuit model of \panelref{fig:intro}{d}. The configurations differ in whether the device nodes are grounded at the PCB socket or at the room-temperature breakout panel. These measurements are illustrative of the contributions of the different spurious RF paths to ground that are to be taken into consideration in the case of large-area devices.

When the device under test (DUT) is not connected (${\mathrm{N_5}}$ open), a single resonance is observed, which originates from the series combination of the bias-tee capacitor $C_{\mathrm{BT}}$ and the resonator inductor $L_{\mathrm{r}}$ (orange trace). Once the DUT is connected, the measured $|S_{\mathrm{11}}|$ response depends strongly on the location at which the device nodes are grounded. When the reference ground is applied at the room-temperature breakout panel, i.e. at the end of the DC loom, additional features appear in the measured $|S_{\mathrm{11}}|$ spectrum that are not reproduced by a model including only PCB and device elements (purple trace). These features are attributed to parasitic contributions from the DC loom wiring. Their presence indicates that RF signals applied at the gate capacitively couple to the source and drain terminals and can propagate along the DC wiring, where additional reactive elements modify the resonant response. In contrast, when the device nodes are grounded locally at the PCB DC socket, the $|S_{\mathrm{11}}|$ response changes markedly and is well described by the circuit model including only PCB and device parasitics (green trace).\\\indent
The configuration used for gate-based reflectometry in this work adopts PCB grounding of the source and drain nodes while maintaining DC access to the gate via a bias-tee. This choice minimises spurious contributions from the DC loom while preserving sensitivity to gate-dependent device impedance.

\section{Effects of temperature on DC and RF characterisation}
\label{app:TEMP}

\begin{figure}[!b]
  \centering
  \includegraphics[width=\columnwidth]{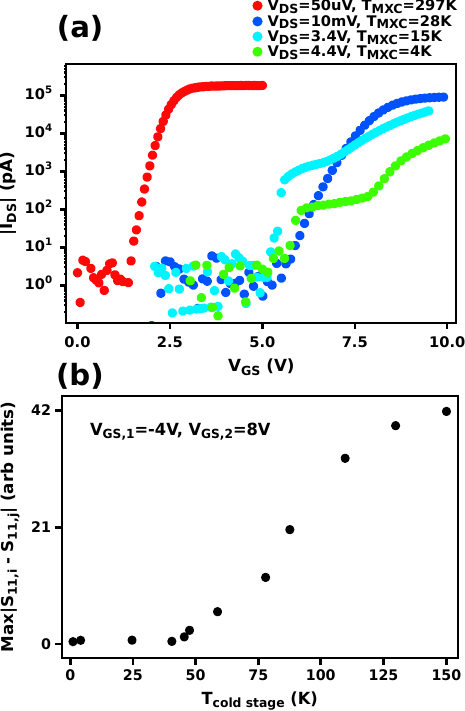}
  \caption{(a) DC transfer characteristics measured at different mixing-chamber temperatures and drain--source biases, showing gate-controlled turn-on across the full temperature range. (b) Maximum change in the reflection coefficient magnitude, $\mathrm{Max} |S_{11,i} - S_{11,j}|$ over a frequency range around $f_0$, as a function of temperature.}
  \label{fig:trnsfrandS11TEMP}
\end{figure}

In Fig.~\ref{fig:trnsfrandS11TEMP}(a), we show the transfer characteristic over a wide temperature range confirming that the device remains functional in DC operation down to 4~K. In contrast, the reflectometry response is progressively suppressed with decreasing temperature and is no longer detectable below $\approx 45$~K, as shown in Fig.~\ref{fig:trnsfrandS11TEMP}(b).

The data in Fig.~\ref{fig:trnsfrandS11TEMP}(b) were acquired while the refrigerator was warming up. The reflection coefficient $S_{\mathrm{11}}$ was measured repeatedly over time, and each trace was assigned a temperature using the cold-stage thermometer at the corresponding time. Although the cold-stage thermometer does not measure exactly the same temperature as the mixing chamber plate to which the device is attached, it is used because it functions over a wider temperature range, enabling measurements to be made both in regimes where an RF response is observed and where it is absent. Any temperature difference between the cold stage and the mixing chamber is expected to be small and not sufficient to affect the qualitative validity of the results.

\section{Estimates of device resistive components}
\label{app:drift_freezeout}

The on-state resistance,
$R_{\mathrm{DS,on}}$, can be decomposed into resistive sub-components contributing differently at room temperature and at $T=28~\mathrm{K}$. 
At room temperature, the source and drain contacts are assumed to be
ohmic and their contribution can be neglected. The measured
room-temperature on-state resistance is therefore written as $R_{\mathrm{DS,on}}^{\mathrm{RT}}
\approx
R_{\mathrm{ch}}^{\mathrm{RT}}
+
R_{\mathrm{drift}}^{\mathrm{RT}}
$, where $R_{\mathrm{ch}}^{\mathrm{RT}}$ is the channel resistance and
$R_{\mathrm{drift}}^{\mathrm{RT}}$ denotes the combined JFET and
drift-region resistance~\cite{baliga}. At cryogenic temperature carrier freeze-out can increase the drift-region resistance, and the source and drain contacts may also contribute appreciable series resistance. We, therefore, write~$R_{\mathrm{DS,on}}^{\mathrm{cryo}}
\approx
R_{\mathrm{ch}}^{\mathrm{cryo}}
+
R_{\mathrm{drift}}^{\mathrm{cryo}}
+
R_{\mathrm{c,S}}
+
R_{\mathrm{c,D}}$, where $R_{\mathrm{c,S}}$ and $R_{\mathrm{c,D}}$ are effective source-
and drain-side contact/access resistances. As discussed in the main text,
at $T=28~\mathrm{K}$ we estimated 
$R_{\mathrm{DS,on}}^{\mathrm{cryo}}\approx
110~k\Omega
$~from DC transfer measurements.

In the lumped element model used to explain our results at cryogenic temperature, we have considered that the increase in drift resistance is the dominant resistive contribution and neglected the other resistive components, as far as the RF response is concerned.
In the following sub-sections, we estimate the drift-region cryogenic resistance from donor
freeze-out and then discuss why the contact resistance, although relevant
for DC transport, is not expected to contribute a comparable resistive
impedance in the RF model. As for the $R_{\mathrm{ch}}$ contribution, although it is also expected to vary with temperature, channel mobility considerations would suggest that this is
a negligible effect compared to the drift resistance~\cite{Das2022}. 

\subsection{Drift-region resistance}

Charge neutrality with incomplete donor ionisation gives
\begin{equation}
n = N_D^+,
\end{equation}
where $n$ is the free electron density, $N_D$ is the donor concentration, and $N_D^+$ is the ionised donor density. The carrier density is given by 
\begin{equation}
n(T) = \frac{-1 + \sqrt{1 + 4\alpha(T)N_D}}{2\alpha(T)},
\end{equation}
where $T$ is the temperature and $\alpha(T)$ is defined as
\begin{equation}
\alpha(T) = \frac{g_D}{N_C(T)} \exp\!\left(\frac{\Delta E_D}{k_B T}\right).
\end{equation}
Here $g_D$ is the donor degeneracy factor, $\Delta E_D$ is the effective donor activation energy, $k_B$ is Boltzmann's constant, and $N_C(T)$ is the effective density of states in the conduction band, given by
\begin{equation}
N_C(T) = N_C(300)\left(\frac{T}{300}\right)^{3/2}.
\end{equation}
For 4H-SiC, we take $N_C(300) \approx 1.7 \times 10^{19}~\mathrm{cm^{-3}}$ \cite{KimotoCooper2014}, $g_D = 2$, and $N_D = 8 \times 10^{15}~\mathrm{cm^{-3}}$ \cite{Liu2020LowRon1200V}. For low-doped $n$-type 4H-SiC, nitrogen donors have been reported at $62~\mathrm{meV}$ and $110~\mathrm{meV}$ with equal donor populations \cite{Robert2001DonorIonis}. In the present appendix, these are represented by a single effective donor ionisation energy of $\Delta E_D = 86~\mathrm{meV}$ for an order-of-magnitude estimate. Evaluating these expressions at $T = 300~\mathrm{K}$ gives
$n(300~\mathrm{K}) \approx 7.8 \times 10^{15}~\mathrm{cm^{-3}}$, while at $T = 28~\mathrm{K}$ the carrier density is strongly reduced by donor freeze-out, giving~$n(28~\mathrm{K}) \approx 8.0 \times 10^{8}~\mathrm{cm^{-3}}$. The corresponding reduction factor is therefore$\frac{n(300~\mathrm{K})}{n(28~\mathrm{K})} \approx 9.7 \times 10^{6}$.\\
The resistivity is given by~$\rho = \frac{1}{q \mu_n n}$, where $q$ is the elementary charge, and $\mu_n$ is the electron mobility. Using $\mu_n(300~\mathrm{K}) \approx 500~\mathrm{cm^2\,V^{-1}\,s^{-1}}$ \cite{KimotoCooper2014} gives
$\rho(300~\mathrm{K}) \approx 1.6~\Omega\,\mathrm{cm}$. Low-temperature measurements on high-quality 4H-SiC show that the electron mobility can increase substantially at cryogenic temperature \cite{Pernot2000LowTMobility}. To reflect this, we take an illustrative value
$
\mu_n(28~\mathrm{K}) \approx 10\,\mu_n(300~\mathrm{K}) \approx 5.0 \times 10^3~\mathrm{cm^2\,V^{-1}\,s^{-1}}.
$
Even with this increase, the resistivity remains dominated by donor freeze-out because the carrier density is reduced by approximately $9.7 \times 10^{6}$. The resulting resistivity at $28~\mathrm{K}$ is $\rho(28~\mathrm{K}) \approx 1.56 \times 10^{6}~\Omega\,\mathrm{cm}$.
\\\indent The drift-region resistance is estimated from
\begin{equation}
R_{\mathrm{drift}} = \rho \frac{t_{\mathrm{drift}}}{A},
\end{equation}
where $R_{\mathrm{drift}}$ is the drift-region resistance, $t_{\mathrm{drift}}$ is the drift-layer thickness, and $A$ is the effective current-carrying area. Taking $t_{\mathrm{drift}} = 10~\mu\mathrm{m}$ \cite{Liu2020LowRon1200V} and $A \approx 18~\mathrm{mm^2}$ gives~$R^\textup{RT}_{\mathrm{drift}} \approx 9~\mathrm{m}\Omega$ at $T=300~\mathrm{K}$
and $R^\textup{cryo}_{\mathrm{drift}} \approx 8.6~\mathrm{k}\Omega$ at $T=28~\mathrm{K}$.
These arguments explain the origin of the numerical values attributed to the drift resistance in the circuit simulations discussed in the main text.

\subsection{Contact impedance in the RF model}

The estimate above shows that the large increase in the drift region resistance at cryogenic temperature cannot alone explain the even larger increase in ON-state resistance ($R_{\mathrm{DS,on}}^{\mathrm{cryo}}\approx 110k\Omega$). Hence, we argue that also contact resistances must increase dramatically. Next, we explain why the contact contributions can be neglected in the RF model.\\\indent
In our RF measurements the source and drain metals are both
locally grounded, so the contacts are probed as small-signal impedances
about approximately zero drain-source bias. If a cryogenic contact is
treated as a Schottky-like barrier, its small-signal impedance can be
represented by a differential resistance in parallel with a depletion
capacitance \cite{SzeNg2007},
\begin{equation}
    Z_{\mathrm{c},x}(\omega)
\approx
\left(
\frac{1}{R_{\mathrm{c},x}}
+
j\omega C_{\mathrm{c},x}
\right)^{-1}.
\end{equation}
Here, \(x\) denotes either the source or drain contact,
\(\omega = 2\pi f\), \(R_{\mathrm{c},x}\) is the zero-bias differential
contact resistance, and \(C_{\mathrm{c},x}\) is the corresponding
contact capacitance. Thus, \(R_{\mathrm{c},x}\) may contribute to
\(R_{\mathrm{DS,on}}^{\mathrm{cryo}}\) in DC, while \(C_{\mathrm{c},x}\)
can bypass this resistance at RF.

The zero-bias contact capacitance is estimated using the depletion
capacitance of a one-sided Schottky-like junction,~$C_{\mathrm{c},x}
\approx
A_{\mathrm{c},x}
\sqrt{
\frac{q\epsilon_{\mathrm{SiC}}N_{\mathrm{eff},x}}
{2V_{\mathrm{bi},x}}
}$. Here, \(A_{\mathrm{c},x}\) is the effective contact area, \(q\) is the
elementary charge, \(\epsilon_{\mathrm{SiC}}\) is the permittivity of
SiC, \(N_{\mathrm{eff},x}\) is the effective ionised dopant density in
the depleted semiconductor region adjacent to the contact, and
\(V_{\mathrm{bi},x}\) is the built-in potential. The source and drain
metals in vertical SiC MOSFETs typically contact highly doped regions,
so \(N_{\mathrm{eff},x}\) is not expected to freeze out in the same way
as the lightly doped drift region; in highly doped 4H-SiC, the donor
activation energy becomes doping concentration dependent and hopping or impurity-band
conduction can contribute \cite{Evwaraye1998}.

For a conservative estimate, we take
\(N_{\mathrm{eff},x}\approx10^{16}~\mathrm{cm^{-3}}\),
\(V_{\mathrm{bi},x}\approx1~\mathrm{V}\),
\(\epsilon_{\mathrm{SiC}}\approx9.7\epsilon_0\), and
\(A_{\mathrm{c},x}\approx18~\mathrm{mm^2}\), comparable to the active
device area. This gives~$
C_{\mathrm{c},x}
\approx
4.5~\mathrm{nF}$. At \(f_0\approx566~\mathrm{MHz}\), the corresponding capacitive
impedance is~$|Z_C|
\approx
\frac{1}{2\pi f_0 C_{\mathrm{c},x}}
\approx
0.06~\Omega $. As such, although contact
resistance is expected to contribute to the measured
\(R_{\mathrm{DS,on}}^{\mathrm{cryo}}\) in DC, source and
drain-contact resistances are not expected to make a significant
contribution to the RF response. Hence, these are neglected in the equivalent circuit model.\\






\section{Gate-based vs Ohmic-based reflectometry}
\label{app:gate_vs_ohmic}
\begin{figure}[b]
  \centering
  \includegraphics[width=\columnwidth]{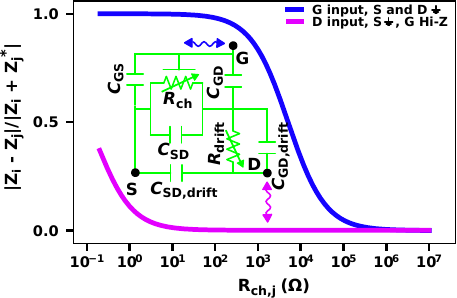}
  \caption{Normalised impedance change in the DUT,
as a function of the channel resistance $R_\mathrm{ch,j}$ at a fixed frequency $f_0 =566~$MHz. The reference impedance, $Z_\textup{i}$, is calculated for a fixed $R_\mathrm{ch,i}=10~\mathrm{M}\Omega$, to represent channel pinch-off. Two configurations are shown: RF signal applied to the gate with source and drain grounded (blue), and RF signal applied to the drain with source grounded and gate at high impedance (Hi-Z, pink). In both cases, $R_{\mathrm{drift}}=8.6~\mathrm{k}\Omega$. The values of the other components used in the simulations are listed in Table~I.}
  \label{fig:ZGDS}
\end{figure}
Reflectometry readout is commonly approached in two alternative ways, i.e. ohmic-based or gate-based~\cite{Vigneau2022APR}. It is important to clarify that in this work we chose gate-based readout because of specific requirements arising from the large-area device architecture.\\\indent
In Fig.~\ref{fig:ZGDS}, we show the DUT normalised impedance change resulting from a variation in $R_\mathrm{ch}$. This is evaluated for the two alternative reflectometry measurement configurations: RF signal applied to the gate with the source and drain grounded (our experimental case), and RF signal applied to the drain with the gate connected to a high impedance DC line and source grounded. In both cases, we fixed $R_{\mathrm{drift}} = 8.6~\mathrm{k}\Omega$, to represent the cryogenic regime. The simulations show that the impedance change is larger for the gate-based configuration over the full range of $R_{\mathrm{ch}}$ chosen to represent the transistor being driven between channel pinch-off and full inversion. Since the impedance modification is directly proportional to the maximum reflection-coefficient change in an ideal (lossless) matching network, the gate-based approach offers (in principle) a more favourable configuration for attaining reflectometry readout~\cite{Vigneau2022APR,Allen2003RFsignalTheory}.\\\indent 
To roughly understand the reason for this discrepancy, one should consider that if the signal is applied to an ohmic contact, it is routed to the other ohmic ground via a large drift capacitance which bypasses the transistor channel entirely. By contrast, if the signal is applied to the gate electrode, a  larger fraction is routed through the channel, hence the larger response seen in Fig.~\ref{fig:ZGDS}. 

\bibliography{biblio}

\end{document}